\documentstyle[aps,prd,epsf,floats]{revtex}
\def\cR{{\cal R}}
\def\cH{{\cal H}}
\begin{document}
\preprint{astro-ph/0101406}
\draft
\twocolumn[\hsize\textwidth\columnwidth\hsize\csname
@twocolumnfalse\endcsname

\title{Enhancement of superhorizon scale
inflationary curvature perturbations}

\author{Samuel M.~Leach$^{1}$, Misao Sasaki$^{2}$, David
  Wands$^{3}$ and Andrew R.~Liddle$^{1}$}

\address{$^1$Astronomy Centre, University of Sussex, Brighton, BN1
  9QJ, United Kingdom}

\address{$^2$Department of Earth and Space Science, Graduate School
  of Science, Osaka University, Toyonaka 560-0043, Japan}

\address{$^3$Relativity and Cosmology Group, School of Computer
  Science and Mathematics,\\University of Portsmouth, Portsmouth PO1
  2EG, United Kingdom}

\date{\today}

\maketitle

\begin{abstract}
    We show that there exists a simple mechanism which can enhance the
  amplitude of curvature perturbations on superhorizon scales during
  inflation, relative to their amplitude at horizon crossing.  The
  enhancement may occur even in a single-field inflaton model, and
  occurs if the quantity $a\dot\phi/H$ becomes sufficiently small, as
  compared to its value at horizon crossing, for some time interval
  during inflation.  We give a criterion for this enhancement in
  general single-field inflation models.
\end{abstract}

\pacs{PACS numbers: 98.80.Cq \hfill OU-TAP-155, PU-RCG 01/04,
  astro-ph/0101406}

\vskip2pc]

\section{Introduction}

The standard, single-field, slow-roll inflation model predicts that
the curvature perturbation on comoving hypersurfaces, ${\cal R}_c$,
remains constant from soon after the scale crosses the Hubble horizon,
giving the formula \cite{Muk,Sas}
\begin{eqnarray}
\label{standard}
 {\cal R}_c\approx{\cal R}_c(t_k)\approx \left(H^2 \over 2\pi\dot\phi
 \right)_{k=aH}
\end{eqnarray}
where $H$ is the Hubble parameter, $\dot\phi$ is the time derivative
of the inflaton field $\phi$, and $t_k$ is a time 
shortly after horizon crossing.
However, one may consider a model in
which slow-roll is violated during inflation.  Recently, Leach \&
Liddle \cite{LeaLid} studied the behavior of the curvature
perturbation in a model in which inflation is temporarily suspended,
finding a large amplification of the curvature perturbation relative
to its value at horizon crossing for a range of scales extending
significantly beyond the Hubble horizon.

In this short paper, we consider single-field inflation models and
analyze the general behavior of the
curvature perturbation on superhorizon scales.
We show analytically when and how this large enhancement occurs.
We find that a necessary condition is that the quantity
$z \equiv a\dot\phi/H$ becomes smaller than its value at the time of horizon
crossing. We then present a couple of integrals which involve
the above quantity and which give a criterion for enhancement.

\section{Enhancement of the curvature perturbation}

We assume a background metric of the form
\begin{eqnarray}
 \label{backgr}
ds^2 &=& -dt^2 +a^2(t)\delta_{ij}dx^i dx^j \nonumber \\
&=& a^2(\eta)\left(-d\eta^2+\delta_{ij}dx^i dx^j\right) \,.
\end{eqnarray}
On this background the growing mode solution of the curvature
perturbation on 
comoving hypersurfaces is known to stay
constant in time on superhorizon scales in the absence of any entropy
perturbation \cite{Muk,Sas,KodSas,MFB,SteLyth}. This follows from the
equation for $\cR_c$
\begin{eqnarray}
  \label{calReq}
 \cR_c''+2{z'\over z}\cR_c'+k^2\cR_c=0,
\end{eqnarray}
where the prime denotes the conformal time derivative, $d/d\eta$, and
$z=a \dot\phi/H$.  One readily sees that on superhorizon scales, when
the last term can be neglected, there exists a solution with $\cR_c$
constant, which corresponds to the growing adiabatic mode.

However, this does not necessarily mean that $\cR_c$ must stay
constant in time after its scale crosses the Hubble horizon.
In fact, if the contribution of the other independent
mode (i.e.~the decaying mode) to $\cR_c$ is large at
horizon crossing, $\cR_c$ will not become constant until the 
decaying mode dies out. The important point here is that
the decaying mode is, by definition, the mode that decays
asymptotically in the future, but it does not necessarily start to
decay right after horizon crossing.
In what follows, we show that there indeed exists a situation
in which the decaying mode can stay almost constant
for a while after the horizon crossing before it starts to decay.
In such a case, the contribution of the two modes to the curvature
perturbation is found to almost cancel at horizon
crossing.
This gives a small initial amplitude of $\cR_c$, but results in a
large final amplitude for $\cR_c$ after the decaying mode becomes
negligible. 

Let $u(\eta)$ be a solution of Eq.~(\ref{calReq}) for any given $k$.
For much of the following discussion it is not necessary to specify
the nature of the solution $u$, but for clarity let us identify it
straightaway as the late-time asymptotic solution at $\eta_*$ (taking
$\eta_*$ for instance as the end of inflation).  For any other
solution, $v(\eta)$, independent of $u(\eta)$, it is easy to show from
Eq.~(\ref{calReq}) that the Wronskian $W=v'u-u'v$ obeys
\begin{equation}
W' = -2 {z'\over z} W \,,
\end{equation}
and hence $W\propto 1/z^2$.
Therefore we have
\begin{eqnarray}
  \left({v\over u}\right)'={W\over u^2}\propto {1\over z^2 u^2}\,.
\end{eqnarray}
Hence the decaying mode, $v$, which vanishes as $\eta\to\eta_*$,
may be expressed in terms of the growing mode, $u$, as
\begin{eqnarray*}
  v(\eta)\propto u(\eta)
\int_{\eta_{*}}^\eta{d\eta'\over
z^2(\eta')u^2(\eta')}\,.
\end{eqnarray*}
Without loss of generality, we may assume $v=u$ at some initial epoch,
which we take to be shortly after horizon crossing, $\eta=\eta_k$
($<\eta_*$). Then $v$ is expressed as
\begin{eqnarray}
  v(\eta)&=&u(\eta){D(\eta)\over D(\eta_k)}\,,
\label{uasv}
\end{eqnarray}
where
\begin{eqnarray}
D(\eta)
=3\cH_k \int_\eta^{\eta_{*}}d\eta'{z^2(\eta_k)u^2(\eta_k)\over
z^2(\eta')u^2(\eta')}\,,
  \label{Ddef}
\end{eqnarray}
and, for convenience, the conformal Hubble parameter
$\cH_k=(a'/a)_k$ at $\eta=\eta_k$ is inserted to make $D$
dimensionless.
In terms of $u$ and $v$, the general solution of $\cR_c$ may be
expressed as 
\begin{eqnarray}
  \label{gensol}
  \cR_c(\eta)=\alpha u(\eta)+\beta v(\eta),
\end{eqnarray}
where $\alpha$ and $\beta$ are constants and we assume
$\alpha+\beta=1$ without loss of generality. 
Thus, if the amplitude of $\cR_c$ at horizon crossing differs
significantly from that of the growing mode, $\alpha u(\eta_k)$, it
can only be because $|\beta|\gg1$.

Using Eq.~(\ref{uasv}) and noting $\alpha+\beta=1$,
$\cR_c$ and $\cR_c'$ at the initial epoch $\eta=\eta_k$ are given by
\begin{eqnarray}
&&\cR_c(\eta_k)=u(\eta_k)\,,
\nonumber\\
&&\cR_c'(\eta_k)=u'(\eta_k)
 -{3 (1-\alpha) \cH_k u(\eta_k)\over D_k}\,,
  \label{iniR}
\end{eqnarray}
where $D_k=D(\eta_k)$.
Then $\alpha$ can be expressed in terms of the initial conditions as
\begin{eqnarray}
\label{defalpha}
  \alpha=1+ D_k{1\over3{\cal H}_k}\left[{\cR_c'\over\cR_c}
-{u'\over u}\right]_{\eta=\eta_k}\,.
\end{eqnarray}
If we assume $\cR_c(\eta_k)$ to be a complex amplitude determined by
an initial vacuum state for quantum fluctuations, then
$\cR_c'/(\cH_k\cR_c)$ at the time of horizon crossing
will be at most of order unity. This implies that $|\alpha|$, and hence 
$|\beta|$, can become large if $D_k\gg1$ or 
$(D_k/\cH_k)|u'/u|\gg1$. 

\section{Long-wavelength approximation}

Equation~(\ref{calReq}) can be written in terms of the canonical field
perturbation, $Q=z\cR$, as
\begin{equation}
\label{uequation}
Q'' + \left(k^2 - {z''\over z} \right) Q = 0 \,.
\end{equation}
{}From this we see that the general solution for $k^2\ll |z''/z|$ is given
approximately by
\begin{equation}
\label{longwavelengthRc}
\cR_c \approx A + B \int_{\eta_*}^\eta {d\eta' \over z^2(\eta')} \,.
\end{equation}
where $A$ and $B$ are constants.

The requirement that $v\to0$ as $\eta\to\eta_*$ uniquely identifies
the decaying mode as proportional to
$\int_\eta^{\eta_*}d\eta'/z^2(\eta')$ in Eq.~(\ref{longwavelengthRc}),
but one is always free to include arbitrary contributions 
from the decaying mode in the growing mode. Nonetheless, it is
convenient to identify the constant $A$ in
Eq.~(\ref{longwavelengthRc}) as an approximate
solution for the growing mode, $u$, on sufficiently large scales.
Thus we put the lowest order solutions for $u$ and $v$ as
\begin{eqnarray}
u_0={\rm const.}\,,\quad v_0=u_0{D(\eta)\over D_k}\,,
\label{lowest}
\end{eqnarray}
where and in the rest of the paper
$D(\eta)$ is the integral given by Eq.~(\ref{Ddef})
but with $u$ approximated by $u_0$,
\begin{eqnarray}
  \label{approxD}
  D(\eta)\approx 3{\cal H}_k\int_{\eta}^{\eta_*}d\eta'
{z^2(\eta_k)\over z^2(\eta')}\,.
\end{eqnarray}

As long as the slow-roll condition is satisfied, the above approximate
solutions are accurate enough. However, the next order correction
to the growing mode $u$ may become substantial if there is an epoch
at which the slow-roll condition is 
violated\cite{StarComm}. It then becomes important
to make clear how one chooses the growing mode.

Rewriting Eq.~(\ref{calReq}) in an iterative form,
with the lowest order solution
$u=u_0$, a growing mode solution is given by
\begin{eqnarray}
&&u=\left[1+F(\eta)\right]u_0\,;
\nonumber\\
&&\quad 
F(\eta)=k^2\int_{\eta}^{\eta_*}{d\eta'\over z^2(\eta')}
\int^{\eta'}_{\eta_k}z^2(\eta''){u(\eta'')\over u_0}d\eta''\,,
\label{Fdef}
\end{eqnarray}
where the boundary condition is chosen so that $u\to u_0$ as $\eta\to\eta_*$.
The above equation tells us when and how the approximation $u\approx u_0$ 
on superhorizon scales may be invalidated once
the $O(k^2)$ effect is taken into account.

In the long-wavelength approximation, $F$ can be
approximated by setting $u=u_0$ in the integral to obtain the solution to
$O(k^2)$ accuracy. Thus we take the approximation,
\begin{eqnarray}
F(\eta)\approx k^2\int_{\eta}^{\eta_*}{d\eta'\over z^2(\eta')}
\int^{\eta'}_{\eta_k}z^2(\eta'')d\eta''\,.
\label{approxF}
\end{eqnarray}
The $O(k^2)$ effect cannot be neglected if this integral
becomes larger than unity. As may be guessed
from the form of the integral,
such a situation appears if there is an epoch during
which $z^2(\eta)\ll z^2(\eta_k)$.

To be specific, let us assume $z(\eta)\ll z_k=z(\eta_k)$ for
$\eta>\eta_0 (>\eta_k)$.
 Then $F(\eta)$ will become large and approximately
constant for $\eta_k<\eta<\eta_0$ and will decay when $\eta>\eta_0$.
Incidentally, this behavior is quite similar to the behavior of
the lowest order decaying mode $v_0(\eta)$ given in Eq.~(\ref{lowest}).
In other words, the growing mode can be substantially contaminated
by a component that behaves like the decaying mode,
and it can no longer be assumed as being constant
on large scales.

The above discussion suggests that we may take advantage
of the ambiguity in defining the growing mode to redefine 
$u$ accurate to $O(k^2)$ by
\begin{eqnarray}
u= \left[{1+F(\eta)}\right]u_0-F_k\,v_0(\eta)\,,
\end{eqnarray}
where $F_k=F(\eta_k)$. The growing mode will be now
approximately constant on superhorizon scales: $u\approx u_0$,
or at least $u(\eta_k)=u(\eta_*)$.
However, $u'/u$ at $\eta=\eta_k$ will no longer be negligible.
We find
\begin{eqnarray}
\left[{u'\over u}\right]_{\eta=\eta_k}
=-F_k\,\left[{v'_0\over v_0}\right]_{\eta=\eta_k}
={3\cH_k\over D_k}F_k\,.
\end{eqnarray}
Then Eq.~(\ref{defalpha}) for $\alpha$ may be
approximated as
\begin{eqnarray}
\alpha\approx1+{D_k\over3\cH_k}{\cR_c'\over\cR_c}-F_k\,,
\label{approxalpha}
\end{eqnarray}
where $D_k$ and $F_k$ are those given in the
long-wavelength approximation, Eqs.~(\ref{approxD})
and (\ref{approxF}),
and for definiteness we will take $(k/\cH_k)^2=0.1$.

In slow-roll inflation, the time variation of $\dot\phi$ is small and
$z$ increases rapidly, approximately proportional to the scale factor
$a$.  Hence neither the integral $D_k$ nor $F_k$ cannot become large.
Soon after horizon crossing $\cR_c'/\cR_c\ll \cH$, 
so that $\alpha\approx1$ and the standard
result $\cR_c(\eta)\approx\cR_c(\eta_k)$ holds. 
However, if the slow-roll condition is violated, $\dot\phi$ may become
very small and $z$ may decrease substantially to give a large value of
$D_k$ and $F_k$.  (The case where $z$ actually crosses zero is treated
separately in an Appendix.) Then at late times, we have
\begin{eqnarray}
  \label{Rinf}
  \cR_c(\eta_*)= \alpha u(\eta_*) \approx \alpha u(\eta_k)
=\alpha \cR_c(\eta_k)\,.
\end{eqnarray}
Thus the final amplitude will be enhanced by a factor
$|\alpha|$, which can be large if $D_k\gg1$ or $F_k\gg1$.

\section{Starobinsky's model}

As an example we consider the model discussed by Starobinsky
\cite{Starobinsky92}, where the potential has a sudden change in its
slope at $\phi=\phi_0$ such that
\begin{equation} 
V(\phi) = \left\{ 
\begin{array}{ll} 
V_0 + A_+ (\phi-\phi_0) \ \ & {\rm for}\ \phi>\phi_0 \\ 
V_0 + A_- (\phi-\phi_0) \ \ & {\rm for}\ \phi<\phi_0
\end{array} 
\right. \,. 
\end{equation}
If the change in the slope is sufficiently abrupt~\cite{Starobinsky92}
then the slow-roll can be violated and for $A_+>A_->0$ the field enters
a friction-dominated transient (or ``fast-roll'') solution with
$\ddot\phi\approx-3H\dot\phi$ \cite{LeaLid} until the slow-roll
conditions are once again satisfied 
\begin{equation} 
3H_0\,\dot\phi = \left\{
\begin{array}{ll} 
-A_+ \ & {\rm for}\ \phi>\phi_0 \\
-A_- -(A_+-A_-)e^{-3H_0\Delta t} \ & {\rm for}\ \phi<\phi_0 
\end{array} 
\right. \,.
\end{equation} 
For $\phi<\phi_0$ we have  
\begin{equation} 
\label{approxz} 
z \simeq -a_0 {A_-e^{H_0\Delta t} + (A_+-A_-)e^{-2H_0\Delta t} \over
  3H_0^2} \,.
\end{equation}
This decreases rapidly to a minimum value $z_{\rm min}\approx
(A_-/A_+)^{2/3}z_0$ for $A_+\gg A_-$, which can cause a significant
change in $\cR_c$ on superhorizon scales.

First let us discuss the behavior of $D(\eta)$.
For a mode that leaves the horizon in the slow-roll regime $z$ grows
proportional to $a$ while $\phi>\phi_0$, so that the integrand of $D(\eta)$
remains small. Hence $D(\eta)\approx D_k$, which implies
$\cR_c(\eta)\approx \cR_c(\eta_k)$ until $\eta=\eta_0$.  Even after
the slow-roll condition is violated $\cR_c(\eta)$ still remains
constant until $z$ becomes smaller than $z_k$ and the
integrand of $D(\eta)$ becomes large again. Then $D(\eta)$ may increase
rapidly until $\cR_c$ approaches the asymptotic value for
$\eta\to\eta_*$, given by Eq.~(\ref{Rinf}).  
Substituting the above solution for $z$ in Eq.~(\ref{approxz}) into
Eq.~(\ref{approxD}) we obtain 
\begin{equation}
  D_k\approx
 \left\{
\begin{array}{ll}
\displaystyle
1 + {A_+\over A_-} \left( {k\over\cH_k} {\cH_0 \over k}
  \right)^3 \ \ & {\rm for}\ k>(k/\cH_k)\cH_0 \\
\displaystyle
1 + {A_+\over A_-} \left( {\cH_k\over k} {k \over \cH_0}
  \right)^3 \ \ & {\rm for}\ k<(k/\cH_k)\cH_0
\end{array}
\right. \,,
\end{equation}
which shows that for $A_+/A_-\gg1$, we have $D_k\gg1$
on scales
$(A_{-}/A_{+})^{1/3}\cH_0\lesssim k \lesssim (A_+/A_-)^{1/3}\cH_0$.

\begin{figure}[t] 
\centering 
\leavevmode\epsfysize=6cm \epsfbox{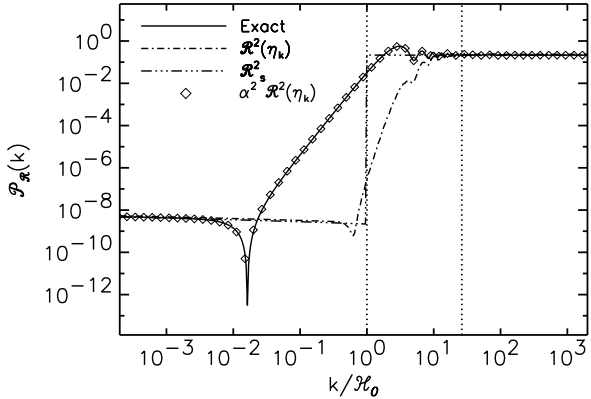}\\
\caption[fig1]{\label{fig:staro_spec1} The power spectrum for the Starobinsky
model~\cite{Starobinsky92} with $A_{+}/A_{-}=10^4$. Plotted are the exact 
asymptotic value of the curvature perturbation $\cR_c^2(\eta_*)$, the
horizon-crossing value $\cR_c^2(\eta_k)$, and the
enhanced horizon-crossing amplitude $\alpha^2\cR_c^2(\eta_k)$
using the long-wavelength approximation. The range of
scales between the dotted lines corresponds to modes leaving the horizon
during the transient epoch, defined as the region where $z'/z<0$. Also
plotted is the slow-roll amplitude $\cR^2_s$ given by Eq.~(\ref{eqn:sr}).}
\end{figure}

A similar behavior is expected for $F(\eta)$. Using again the solution
for $z$ in Eq.~(\ref{approxz}), the double integral in Eq.~(\ref{Fdef})
is evaluated to give
\begin{eqnarray}
F_k\approx
 \left\{
\begin{array}{ll}
\displaystyle
{1\over15}{A_+\over A_-}
\displaystyle
 \left( {k\over\cH_k} {\cH_0 \over k}\right)^5
\ \ & {\rm for}\ k>(k/\cH_k)\cH_0 \\
\displaystyle
{2\over5}{A_+\over A_-}\left({k \over \cH_0}\right)^2
 \ \ & {\rm for}\ k<(k/\cH_k)\cH_0
\end{array}
\right. \,.
\end{eqnarray}
Thus $F_k\gg1$ for
$(A_{-}/A_{+})^{1/2}\cH_0\lesssim k \lesssim(A_{+}/A_{-})^{1/5}\cH_0$.

Combining the effects of $D_k$ and $F_k$, we see that
the correction due to $F_k$ dominates on scales
$k<\cH_0$ and $D_k$ on scales $k>\cH_0$.
In particular the spiky dip in the spectrum seen in Fig.~\ref{fig:staro_spec1}
at $k\sim (A_{-}/A_{+})^{1/2}\cH_0$ is caused by $F_k$, i.e.,
it is the $O(k^2)$ effect in the perturbation equation (\ref{calReq}).
To summarize, the curvature perturbation is significantly affected by
the discontinuity at
$\phi\sim\phi_0$ even on superhorizon scales from
$k\sim (A_{-}/A_{+})^{1/2}\cH_0$ up to $k \sim (A_{+}/A_{-})^{1/3}\cH_0$.

\begin{figure}[t] 
\centering 
\leavevmode\epsfysize=6cm \epsfbox{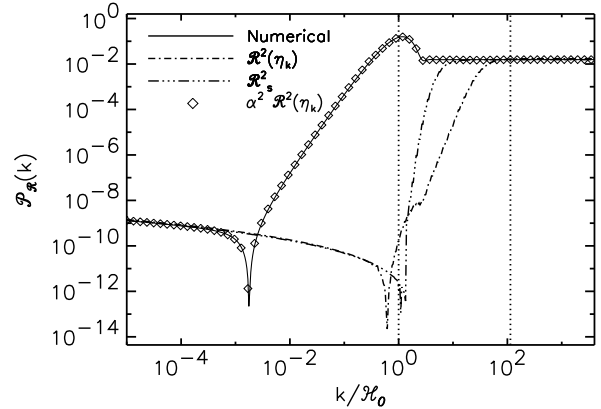}\\
\caption[fig1]{\label{fig:fvq_spec1} Power-spectrum for the false-vacuum
quartic model as in Fig.~\ref{fig:staro_spec1}.}
\end{figure}

Similar behavior was observed in the model studied by Leach \& Liddle
\cite{LeaLid} for false-vacuum inflation with a quartic
self-interaction potential \cite{Roberts}, whose power spectrum is
shown in Fig.~\ref{fig:fvq_spec1}.
In this model there is no discontinuity in the potential,
so the oscillations seen in Starobinsky's model are washed out.  

In both cases our analytic estimate of the enhancement
on superhorizon scales is in excellent agreement with the numerical
results on all scales. Thus our approximate formula for
$\alpha$ given by Eq.~(\ref{approxalpha}) will be very useful
for estimation of the curvature perturbation spectrum
in general models of single-field inflation.

It may be noted that in the Leach \& Liddle model 
the long-wavelength condition, $k^2\ll|z''/z|$,
is violated for modes $k<\cH_0$. It is rather surprising
that our long-wavelength approximation still works very
well for this model.

\section{Invariant spectra}

A striking feature of these results is that the modes which leave the
horizon during the transient regime share the same underlying spectrum
as that produced during the subsequent slow-roll era. This is a
manifestation of the `duality invariance' of perturbation spectra
produced in apparently different inflationary scenarios~\cite{DW}.

Starting from a particular asymptotic background solution, $z(\eta)$,
one finds a two parameter family of solutions
\begin{equation}
\label{generalz}
\tilde{z}(\eta) = C_1 z(\eta) + C_2 z(\eta)\int_\eta^{\eta_*} {d\eta' \over
  z^2(\eta')} 
\,, 
\end{equation}
which leave $z''/z$ unchanged in the perturbation
equation~(\ref{uequation}) and thus generate the same perturbation
spectrum from vacuum fluctuations~\cite{DW} (up to the overall
normalization $C_1$). 
The variable $z$ itself obeys the second-order equation 
\begin{eqnarray}
z''+\left(a^2{d^2V\over d\phi^2}-5\cH^2+\cH'+2{\cH''\over\cH}
-2{\cH'{}^2\over\cH^2}\right)z=0\,.
\end{eqnarray}
Thus for a weakly interacting field
($d^2V/d\phi^2\approx$ constant) in a quasi-de Sitter background
($H\approx$ constant) the equation can be approximated by the linear
equation of motion
\begin{equation}
z'' + \left( a^2{d^2V\over d\phi^2} - 2\cH^2 \right) z \approx 0 \,.
\end{equation}
The general solution $\tilde{z}(\eta)$ is related
to the asymptotic late-time solution $z(\eta)$ by the expression given
in Eq.~(\ref{generalz}).

This means that the usual slow-roll result [taking
$\dot\phi\approx-(dV/d\phi)/3H$] for the amplitude of
the curvature perturbations in Eq.~(\ref{standard})
\begin{equation} 
 \cR_c \approx - \left( {3H^3 \over 2\pi(dV/d\phi)} \right)_{k=\cH}  
 \,, \label{eqn:sr}
\end{equation}
may continue to be a useful approximation even when the actual
background scalar field solution at horizon crossing is no longer
described by slow-roll, as was noted previously by Seto, Yokoyama and
Kodama~\cite{SYK} and seen in our figures.

\section{Summary}

In summary, we have studied the enhancement of the curvature
perturbation on superhorizon scales possible in some models of
inflation. We have found that the curvature perturbation can be
enhanced on superhorizon scales even in single-field inflation,
provided that the slow-roll condition is violated and $a\dot\phi/H$
becomes small compared to its value at horizon crossing. We have
presented a quantitative criterion for this enhancement, namely that
either of the integrals $D_k$ and $F_k$ defined by
Eqs.~(\ref{Ddef}) and (\ref{Fdef}), respectively,
becomes larger than unity. In the long-wavelength
approximation ($k^2\ll |z''/z|$) these integrals are expressed
in terms of the background quantity $z=a\dot\phi/H$, 
as given by Eqs.~(\ref{approxD} and (\ref{approxF}),
so an analytical formula for the final
curvature perturbation amplitude may be derived without assuming
slow-roll inflation. In the case of a weakly self-interacting field in
de Sitter inflation we recover the usual slow-roll formula for the
amplitude of the scalar perturbations even when the background
solution is far from slow-roll at horizon crossing.

\section*{Acknowledgements}

We would like to thank Karim Malik, Alexei Starobinsky and Jun'ichi
Yokoyama for useful discussions and comments.
S.M.L.~is supported by PPARC, M.S.~in part by Yamada Science Foundation 
and D.W.~by the Royal Society. 

\appendix

\section{If $\dot\phi$ crosses zero}

The case when $\dot\phi$ and hence $z$ changes its sign can be treated
as follows.  For simplicity, let us assume $z$ changes the sign only
once at $\eta=\eta_0$. Since the integral $F_k$ is still well-defined
in this case, we focus on the integral $D_k$.

In the vicinity of $\eta=\eta_0$, $z$ can
be expressed as $z=z_0'(\eta-\eta_0)$ where $z_0'=z'(\eta_0)$.
 Hence the equation for ${\cal R}_c$ becomes
\begin{eqnarray}
\left[{d^2\over d\eta^2}+{2\over\eta-\eta_0}{d\over d\eta}+k^2\right]
{\cal R}_c=0.
\end{eqnarray}
The two independent solutions can be found as
\begin{eqnarray}
&&u\approx C\left(1-{1\over6}k^2(\eta-\eta_0)^2+\cdots\right),
\label{growsol}\\
&&v\approx D\left({1\over\eta-\eta_0} -
 {1\over2}k^2(\eta-\eta_0)+\cdots\right). 
\label{decaysol}
\end{eqnarray}
It is apparent that $u$ should be chosen as the growing mode, and it 
remains constant across the epoch $\eta=\eta_0$.

We require $v$ to describe the decaying mode.
As before, we consider an integral expression of $v$
in terms of $z^2$ and $u$.
Then
\begin{eqnarray}
v=u\int_\eta^{\eta_*}{d\eta'\over z^2 u^2}
\approx u\int_\eta^{\eta_*}{d\eta'\over z^2C^2}
\end{eqnarray}
for $\eta>\eta_0$.
This $v$ behaves in the limit $\eta\to\eta_0+0$ as
\begin{eqnarray}
v\sim{1\over z_0'{}^2C^2(\eta-\eta_0)}\,.
\end{eqnarray}
This should be extended to the region $\eta<\eta_0$ 
as the solution (\ref{decaysol}), which implies
\begin{eqnarray}
v=u \lim_{\epsilon\to0}
        \left(\int_\eta^{\eta_0-\epsilon}{d\eta'\over z^2 u^2}
        +\int_{\eta_0+\epsilon}^{\eta_*}{d\eta'\over z^2 u^2}
        -{2\over z_0'{}^2C^2\epsilon}\right),
\end{eqnarray}
for $\eta<\eta_0$.
Thus introducing the function $\tilde D(\eta)$ by
\begin{eqnarray}
\frac{\tilde D(\eta)}{3{\cal H}_k} =
        \lim_{\epsilon\to0}
        \left(\int_\eta^{\eta_0-\epsilon}d\eta'{z_k^2 u_k^2\over z^2 u^2}
        +\int_{\eta_0+\epsilon}^{\eta_*}d\eta'{z_k^2 u_k^2\over z^2 u^2}
        \right. \nonumber\\
 \left.
        -{2\over\epsilon}{z_k^2 u_k^2\over z'_0{}^2 u_0^2}\right),
\end{eqnarray}
and $\tilde D_k=\tilde D(\eta_k)$, 
where 
$u_0=u(\eta_0)$, the decaying mode 
$v$ normalized to $u$ at $\eta=\eta_k$ is given by
\begin{eqnarray}
v(\eta)=u(\eta){\tilde D(\eta)\over \tilde D_k}\,.
\end{eqnarray}
Thus exactly the same argument applies to this case,
by replacing the original $D_k$ by the above $\tilde D_k$.

\end{document}